\documentclass[structabstract]{aa}
\usepackage{graphicx}
\usepackage{color}
\usepackage{txfonts}

\begin{document}
   \title{Frequency drifts of 3-min oscillations in microwave and EUV emission above sunspots}

  \author{R. Sych\inst{1,2}, T. V. Zaqarashvili\inst{3,4}, V. M. Nakariakov\inst{5,6}, S.A. Anfinogentov\inst{2}, K.Shibasaki\inst{7} and Y. Yan\inst{1}}
          
	\institute{National Astronomical Observatory (NAOC), A20 Datun Road, Beijing 100012, China\\
           \email{sych@iszf.irk.ru}
	   \and
           Institute of Solar-Terrestrial Physics, Irkutsk, Lermontov St., 126a, 664033, Russia     
           \and
Space Research Institute, Austrian Academy of Sciences, Schmiedlstrasse 6, 8042 Graz, Austria
  \and
Abastumani Astrophysical Observatory at Ilia State University, Kazbegi ave. 2a, Tbilisi, Georgia
\and
           Centre for Fusion, Space and Astrophysics, Department of Physics, University of Warwick, Coventry CV4 7AL, United Kingdom
 \and
 Central Astronomical Observatory at Pulkovo of the Russian Academy of Sciences, 196140 St Petersburg, Russia     
 \and
 Nobeyama Solar Radio Observatory/NAOJ, Nagano 384-1305, Japan     
            }

   \date{Received: {...}, 2011; accepted *}


\titlerunning{Frequency drifts of 3-minute oscillations in microwave and EUV above sunspots}
\authorrunning{Sych et al.}
\abstract
{}
{
 We analyse 3-min oscillations of microwave and EUV emission generated at different heights of a sunspot
 atmosphere, studying the amplitude and frequency modulation of the oscillations, and its relationship 
 with the variation of the spatial structure of the oscillations.
}
{
 High-resolution data obtained with the Nobeyama Radioheliograph, TRACE and SDO/AIA are analysed
 with the use of the Pixelised Wavelet Filtering and wavelet skeleton techniques.
}
{ 3-min oscillations in sunspots appear in the form of repetitive trains of the duration 8-20 min (13 min in average). The typical interval between the trains is 30-50 min. The oscillation trains are transient in frequency and power. We detected a repetitive frequency drifts of 3-min oscillations during the development of individual trains. Wavelet analysis shows that there are three types of the frequency drift: positive drifts to high frequencies, negative drifts and fluctuations without drift.  Negative drifts were found to occur more often. The start and end of the drifts coincide with the start time and end of the train. The comparative study of 3-min oscillations in the sequences of microwave and EUV images show the appearance of new sources of the oscillations in sunspots during the development of the trains. These structures can be interpreted as waveguides that channel upward propagating waves, responsible for 3-min oscillations. A possible explanation of the observed properties is the operation of two simultaneous factors: dispersive evolution of the upwardly-propagating wave pulses and the non-uniformity of the distribution of the oscillation power over the sunspot umbra
with different wave sources corresponding to different magnetic flux tubes with different physical conditions and line-of-sight angles.}
{}

\maketitle
%

\begin{figure*}
\centering
\includegraphics[width=16cm, height=9cm]{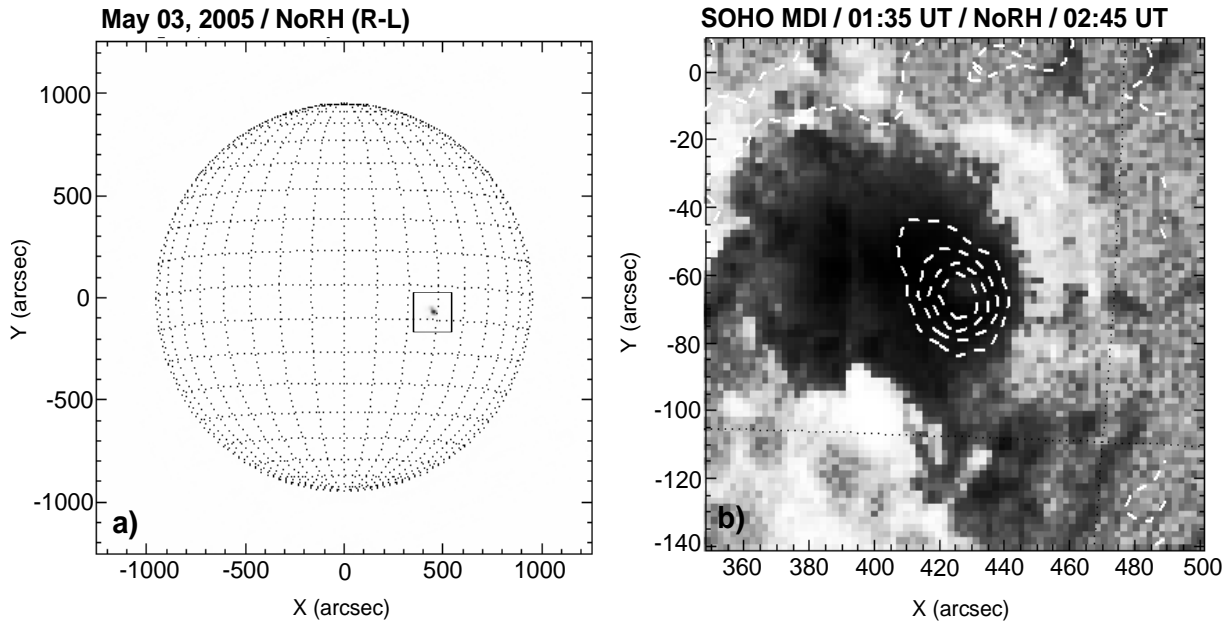}
\caption{Left panel: Full-disk microwave polarisation image of the Sun at 17 GHz on May 03, 2005 at 02:45 UT. 
The rectangle shows the radio source associated with the sunspot. Right panel: Zoomed image of the radio source shown in the left panel; the dashed contours show the L-polarisation image overlapping with the SOHO/MDI magnetogram taken at 01:35 UT. The grey scale indicates the polarity of the magnetic field.}
\label{fig1}
\end{figure*}

\section{Introduction}

Analysis of multi-wavelength data obtained with the new generation of solar observational tools allows us to study dynamics of the solar atmosphere at different heights with high spatial, spectral and temporal resolution. One of the interesting topics is 3-min oscillations in sunspots (e.g. Orrall \cite{Orrall}, Elliott \cite{Elliott}, Liu \cite{Liu}).  Fluctuations in the 3-min range are relatively weak at the photospheric level of sunspots, but they are significantly enhanced in their chromospheres (Lites \cite{Lites1986},
Bard and Carlsson \cite{Bard}, Tziotziou et al. \cite{Tziotziou}, Kobanov et al. \cite{2008AstL...34..133K, 2011A&A...525A..41K}) and in the transition region (Gurman et al. \cite{Gurman},
Lites \cite{Lites1992}, Bogdan \cite{Bogdan}). Observations of the photospheric magnetic field oscillations (Norton et al. \cite{Norton}, Balthasar et al. \cite{Balthasar}, Bellot Rubio et al. \cite{Bellot}) show a fine fragmentation of the magnetic field in the regions where the main power of fluctuations is localised inside sunspots. For example, optical observations show that 5-min oscillations of the magnetic field are
mainly concentrated in the isolated tubes of magnetic field (pores) outside the big sunspots, as well as at the umbra-penumbra boundary (Zhugzhda et al. \cite{Zhugzhda}).
Blinkers or explosive events are also observed above sunspots with a similar time scale, from 400 to 1600 s  (Harrison et al. \cite{Harrison}, Brkovich et al. \cite{Brkovich}).

The first observations of the fine spatial structure of oscillating radio sources above strong sunspots at the transition region level were obtained by Gelfreikh et al.  (\cite{Gelfreikh1999}) with the Nobeyama Radioheliograph.  That study showed that the main source of broadband oscillations is concentrated in the centre of the source of the polarised radio emission at 17 GHz, coincident with the sunspots umbra.  Further investigations in the radio band (Shibasaki et al., 2001) using the correlation curves and images of radio sources showed that 3-min oscillations are present in sunspots during long time.
It was concluded that the brightness fluctuations of the radio sources could be caused by the density and temperature perturbations in acoustic waves propagating through the third gyroresonant level (corresponding to the magnetic field of about 2~kG). 

Nindos  et al. (\cite{Nindos}) used the difference radio maps obtained by VLA at 8.5 and 5 GHz with high spatial resolution. They showed that the amplitude of 3-min oscillations varies 
irregularly in time, with a rapid rise phase, and a rather slow decay phase. The spatial sources of the
amplitude variations are localised in fine patches of a circular shape. 

Sych and Nakariakov \cite{Sych2008} studied the spatial, temporal and phase behaviour of sunspot oscillations, applying the method of Pixelised Wavelet Filtering (PWF). It was shown that the source of 3-min oscillations is located in the centre of the microwave source coinciding with the sunspot umbra. 
On the other hand, 5-min oscillations were found to be mainly located in small, symmetric patches at the the umbra-penumbra boundary, similar to the findings obtained with VLA. Oscillations in different patches were found to have different phases.

Phenomenological relationship between 3-min oscillations of the radio flux from a sunspot and quasi-periodic bursts of energy release in coronal active regions nearby was established by Sych et al. (\cite{Sych2009} and Nakariakov et al. (\cite{Nakariakov}). Recently, Sych et al. (\cite{Sych2010}) used the system for automatic detection of solar oscillations (http://pwf.iszf.irk.ru), that showed that the power of  5-min oscillations in sunspots is concentrated in ring-shaped sources at the upper chromosphere or transition region, which coincide with the footpoints of coronal EUV fans. These spatial details are associated with both the standing waves (symmetric spatial patches) and propagating waves (seen as transient, about 10-30 min, v-shaped increases in the spatial distribution of the oscillation power). The amplitude increase was seen simultaneously in microwaves and in EUV.

Rendtel et al. (\cite{Rendtel}) detected variable changes (drifts) of the periods of sunspot oscillations, using the SOHO/SUMER spectrograph (CIV, NeVIII) and EIT, 171\AA\ . The fluctuations of the Doppler shift
and the emission intensity occur in the same frequency range, 3-7 mHz, with a significant time-drift to higher frequencies. 
Most of the fluctuations in the transition region and lower corona are concentrated near the frequency of 6.2 mHz, and form oscillation trains lasting from 10 to 20 min, sometimes up to 40 min.  
A possible explanation of the drifts is the assumption that the curvature of the magnetic tubes above the sunspot leads to different directions of MHD wave propagation relative to the observer and, accordingly,  to the periodic changes of the Doppler velocity and the sign of the frequency drift. Fludra (\cite{Fludra}) found a significant frequency drift of 5.8 mHz oscillations above sunspots in the transition region lines with SOHO/CDS. The typical duration of the train 
was found to be about 15 minutes. Different frequencies in the spectrum were found to have different spatially separated sources.

Two main mechanisms were suggested to explain 3-min sunspot oscillations: a chromospheric cavity (or resonator) 
and oscillations at the acoustic cut-off frequency. The first mechanism implies that the chromosphere is a resonant cavity for waves, which are reflected from the photosphere due to sharp density gradient and from the transition region due to the sharp temperature gradient (Leibacher and Stein \cite{Leibacher1981}, Leibacher et al. \cite{Leibacher1982}). 
It was recently shown numerically that quasi-monochromatic acoustic oscillations in the 3-min band are a natural response of 
the chromospheric resonator to a broadband (e.g., impulsive) excitation (\cite{2011ApJ...728...84B}).
The second mechanism implies that the 3-min oscillations are resulted from the wake behind propagating disturbances in the stratified atmosphere, which oscillates at the acoustic cut-off frequency (Fleck and Schmitz \cite{Fleck}, Kalkofen et al. \cite{Kalkofen}, Sutmann and Ulmschneider \cite{Sutmann}, Kuridze et al. \cite{Kuridze}). The nonlinear wake behind  propagating acoustic pulses may develop in quasi-periodic shocks, which may shape the plasma properties in the chromosphere and the lower corona. These quasi-periodic shocks have been recently suggested as a mechanism for the quasi-periodic appearance of solar spicules and 5-min oscillations in intensity observed in the solar corona (Murawski and Zaqarashvili \cite{Murawski}, Zaqarashvili et al. \cite{Zaqarashvili}).

In this paper we present results of a comprehensive study of 3-min oscillations
at different heights over sunspots with the use of high-resolution EUV and microwave data obtained 
with SDO/AIA, TRACE  and NoRH.
The aim of the paper is to reveal the fine spatial-temporal structure of 3-min oscillations, 
contributing to the understanding of physical mechanisms for their generation and
developing plasma diagnostics techniques.
In section 2 we describe the instruments and the analysed datasets. Section 3 describes the methods used in the analysis. Section 4 describes the results obtained. Sections 5 and 6 present discussion and conclusions, respectively.

\begin{figure}
\resizebox{\hsize}{!}{\includegraphics[viewport=28 356 535 697]{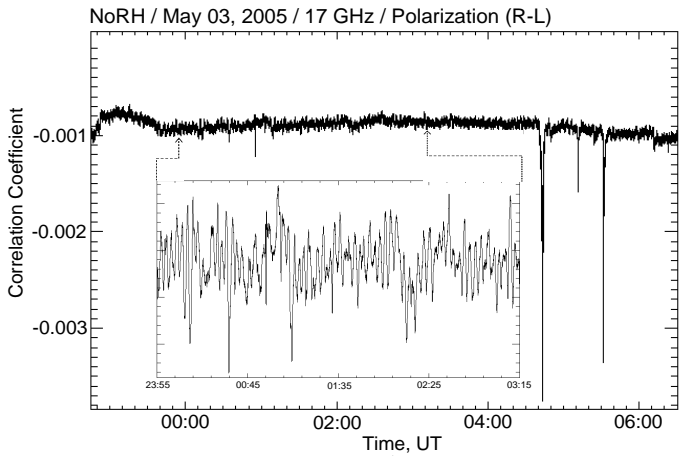}}
\caption{A correlation curve of NoRH in the circular polarisation (R-L) channel. The zoomed time interval,
 selected by the rectangle, shows the smoothed 3-min radio oscillation profile between 23:55 and 03:15 UT. }
\label{fig2}
\end{figure}

\section{Instruments and observations}

To study oscillations in the microwave emission from sunspots we used the Nobeyama Radioheliograph (NoRH, see Nakajima \cite {Nakajima}) data at 17 GHz. Both correlation curves and interferometric images with the spatial pixel size of 4.9 arcsec were analysed. The image synthesis and cleaning of compact sources were made with the Hanaoka algorithm.  The images 
were synthesised every 10 s. We use the data obtained in the circular polarisation channel, where  3-min oscillations are most evident (e.g. Abramenko and Tsvetkov \cite{Abramenko}, Shibasaki \cite{Shibasaki},  Gelfreikh et al. \cite{Gelfreikh1999}). We selected 11 time intervals during 2005-2010 when only one well-developed and isolated sunspot, with a sufficiently strong magnetic field, was present on the solar disk:  January 23-25, 2002 (NOAA 9787), May  02-05, 2005 (NOAA 0756), July 4, 2006 (NOAA 0898), August 15, 2006 (NOAA 0904)  April 29, 2007 (NOAA 0953) and December 8, 2010 (NOAA 11131). Some of these events were discussed in our previous studies (Sych and Nakariakov \cite{Sych2008},  Sych et al. \cite{Sych2009},  Sych et al. \cite{Sych2010}).

Figure~\ref{fig1} shows a typical microwave image, used in the analysis.  The image of the full solar disk was obtained with NoRH at 17 GHz on 3 May 2005 in the circular polarisation channel. There is only one active region, NOAA 10756 (marked by the rectangle in Fig~\ref{fig1}a), with a compact, highly polarised radio source of the L-polarity, with the polarisation degree of about 80\% and the brightness temperature of about 127,000 K. Its spatial coordinates coincide with the large, symmetric sunspot seen in the magnetogram. The strength of the magnetic field was estimated with SOHO/MDI. At 01:35 UT on May 3, 2005 (Fig.~\ref{fig1}b)  the active region with the centre coordinates (430, -60) was mainly unipolar, with prevailing S-polarity (dark areas) and a few isolated islands of the opposite polarity (bright areas). The value of the magnetic field was about 2,700~G.  The polarised radio source at 17 GHz (dashed contour lines), corresponding to the extraordinary electromagnetic wave, synthesised at 02:45 UT, basically coincides with the region of the strong magnetic field.

Series of EV and EUV images on May 4, 2005 and December 8, 2010 were obtained with the Transition Region and Coronal Explorer (TRACE, see Handy and Acton \cite{Handy}) and the Atmospheric Imaging Assembly (AIA) of the Solar
Dynamic Observatory (Title et al. \cite{Title}), respectively. The TRACE data were processed using standard procedures: to remove the background and spike-like brightening due to cosmic rays, to calibrate, and where it was necessary to make normalisation of the exposure time, and the removal of differential rotation during the observation  period. The SDO images were obtained from level 1 data at nine wavelengths: 171, 193, 211, 304, 131, 335, 94, 1600 and 1700\AA\ , with the standard processing such as bad-pixel removal, despiking and flat-fielding and centring the images in a temporary data cube. 

\begin{figure}
\resizebox{\hsize}{!}{\includegraphics[viewport=18 200 541 580]{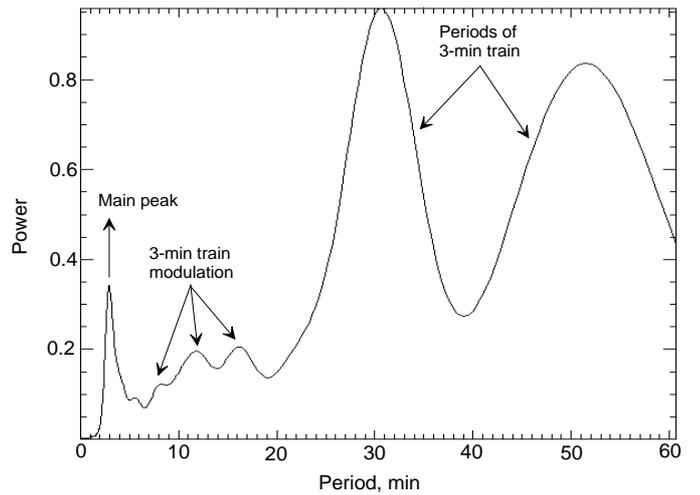}}
\caption{Global wavelet spectrum of the NoRH correlation curve on May 3, 2005. Arrows indicate the main peaks in the spectrum.}
\label{fig3}
\end{figure}

\begin{figure*}
\centering
\includegraphics[width=17cm, height=19cm]{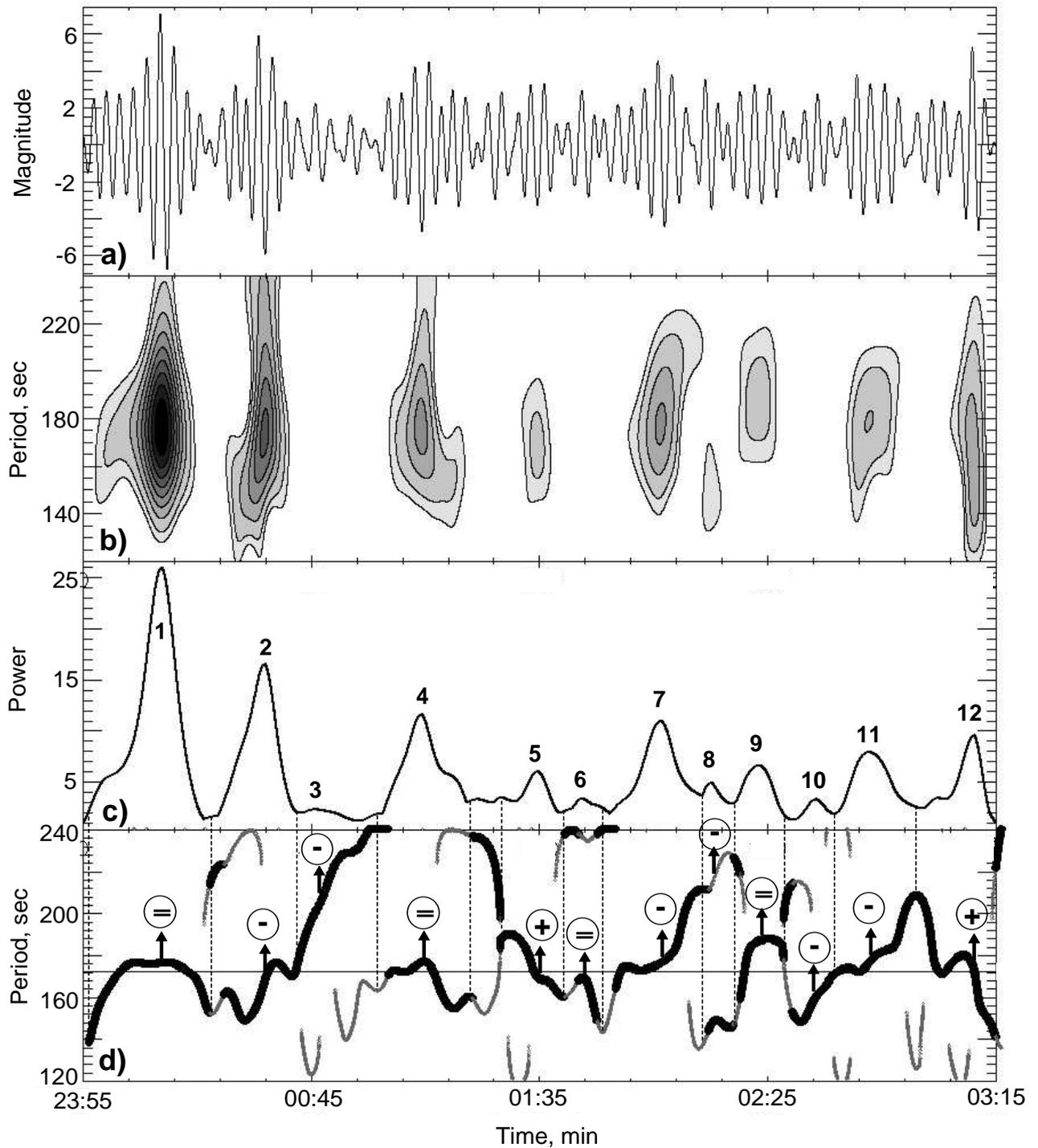}
\caption{Wavelet analysis of the 3-min oscillation of the microwave signal on May 3, 2005 from 23:55 to 03:15 UT.  Panel (a) shows the amplitudes obtained by calculating the inverse wavelet transform of the signal in the 120-240 s period range. Panel (b): power distribution of the 3-min oscillation trains in the wavelet spectrum.  Panel (c): time evolution of the 3-min oscillation power.  The numbers show the numeration of the trains.  Panel (d) shows the wavelet skeleton representing both global (thick lines) and local (thin lines) extrema.  In the circles we show the signs of the period drifts (positive, negative or without drift).  Time is shown in UT, power and amplitudes are in arbitrary units.}
\label{fig4}
\end{figure*}

%
\begin{table*}
\caption{Parameters of 3-min oscillations: the date, the start and end time of the observation, duration of the observation (in minutes), the average period of the spectral power peak, average duration of 3-min oscillation trains,  
times between the trains,  and the fractions of the number of trains with positive, negative and variable drifts.}
\label{table:1}
\centering
\begin{tabular}{ccccccccccclllllllllll}
\hline\hline

No & Date & Start & End & Duration    & Main & Trains & Time & Negative & Positive & Variable \\
  &      &  UT   &  UT & Observation & Peak &  Duration  & Separation &  Drift   &   Drift  &  Drift  \\
  &      & (h,m) &(h,m)&  (min)      & (min)& (min)  &  (min) &  (\%)    &   (\%)   &  (\%)   \\
\hline

1 & 02.05.2005 & 03:05  & 06:25 & 200 & 2.9 & 11.7 & 13,20,40,61    & 53 & 34  & 13\\
2 & 03.05.2005 & 23:55  & 03:17 & 200 &	2.8 &	13.3 &	8,12,16,29,52 &	60 &	32 &  8\\
3 & 04.05.2005 &	22:55 &	01:45 &	170 &	2.9 &	11.3 &	15,30,39,55   &	45 &	51 &	4\\
4 & 05.05.2005 &	00:35 &	03:55 & 200 &	2.6 &	10.2 &	13,15,21,38,51&	56 &	43 &	1\\
5 & 23.01.2002 &	23:59 &	05:00 &	300 &	2.8 &	10.6 &	11,18,30,37,57&	43 &	54 &	3\\
6 & 24.01.2002 &	01:30 &	04:50 &	200 &	3.0 &	11.1 &	11,19,27,46,67&	51 &	47 &	2\\
7 & 25.01.2002 &	23:10 &	02:20 &	190 &	2.8 &	11.8 &	13,23,32,46   &	53 &	47 &	0\\
8 & 04.07.2006 &	00:45 &	05:45 &	300 &	2.8 &	11.6 &	16,33,43,58   &	43 &	51 &	6\\
9 & 15.08.2006 &	23:55 &	04:55 &	300 &	2.8 &	13.2 &	27,36,49,74   &	62 &	36 &	2\\
10& 29.04.2007 &	23:15 &	05:35 &	380 &	2.8 &	11.1 &	20,26,42,54   &	48 &	46 &	9\\

\hline
\end{tabular}
\end{table*}

\section{Methods}

Analysis of transient spatially-varying imaging datasets is a nontrivial task that require the use of dedicated tools.
In our study, we applied the techniques of the wavelet transform with the Morlet mother function (Torrence and Compo \cite{Torrence}) and
the Pixelised Wavelet Filtering (PWF, Sych and Nakariakov \cite{Sych2008}). The PWF 
technique allows us to make narrowband maps of power,
amplitude and phase distribution of oscillations sources at selected harmonics, and obtain their temporal dynamics.  
The data were prepared with the use the web-based interactive system for the remote processing of imaging datasets (http://pwf.iszf.irk.ru, Sych et al. \cite{Sych2010}).

Transient increase in the oscillation power, leading to the local increase in the corresponding wavelet coefficients, was analysed with the method of wavelet skeletons (Mallat \cite{Mallat}). The method is based on extracting only those coefficients in the wavelet decomposition, whose complex amplitudes are highest.  Periodicities in a real signal, decomposed to a set of frequency harmonics of different amplitudes, form local peaks in the wavelet spectrum. Mapping their skeletons extracts the most-powerful harmonics, and track their time and frequency drifts.
One can construct two types of skeletons, considering the behaviour of wavelet coefficient extrema either in the frequency or the time domains. In this study, we use the time skeletons, applying the following algorithm:
 \begin{enumerate}
      \item A power wavelet spectrum with the use of a temporal Morlet mother function is constructed.
      \item The frequency and time domains in the wavelet spectrum, where oscillations take place, are identified.
      \item All local maxima of wavelet coefficients along the time axis in the selected time-frequency region are determined.
      \item The same is repeated for global maxima.
      \item The wavelet skeleton combining the local and global maps is obtained.
  \end{enumerate}

\section{Results}

\subsection{Temporal structure of 3-min microwave oscillations}

Figure~\ref{fig2} shows the profile of the NoRH correlation curve obtained in the  circular polarisation channel (R-L) between 22:45, 3 May 2005 and 06:30 UT, 4 May 2005 with 1-s cadence. The relative amplitude of the oscillations is about 3-10\%, reaching in some peaks 17\%. The 3-min oscillations, clearly seen in the correlation curve,
are mainly related to the variation of the sunspot radio emission. 
Also, the correlation curve contains several spikes, corresponding to irregular energy releases. 
It should be noted that the radio flux from the source changed slightly in comparison to the full Sun flux, which could be attributed to the high level of atmospheric noise. 
The use of  NoRH correlation curves allows one to magnify the signal associated with small-scale sources on the solar
disk, and hence to carry out preliminary analysis of oscillations without synthesising interferometric maps. 
For the detailed analysis, we chose the time interval between 23:55 UT, May 2 and 03:15 UT, May 3, 2005. The zoomed part of this interval, shown inside the rectangle in Figure~\ref{fig2}, clearly demonstrates the presence of well-pronounced 
3-min oscillations. The duration of the interval, 200 min, and the cadence time, 1 s, allowed us to study oscillations with periods from 3 s to 60 min. 

Spectral wavelet transform of the correlation curve gives us its global wavelet spectrum (GWS) shown in Figure~\ref{fig3}.
The GWS spectrum has several well-pronounced peaks, corresponding to the periods of about 3 min and also to 8, 12 and 16 min, and the long-period components of 30 and 50 min. We would like to point out that long-period spectral peaks have already been detected in the emission from sunspots, see, e.g. Gelfreikh et al. (\cite{Gelfreikh1999}) and Chorley et al. (\cite{2010A&A...513A..27C}). 

\begin{figure*}
\centering
\includegraphics[width=18cm, height=11cm]{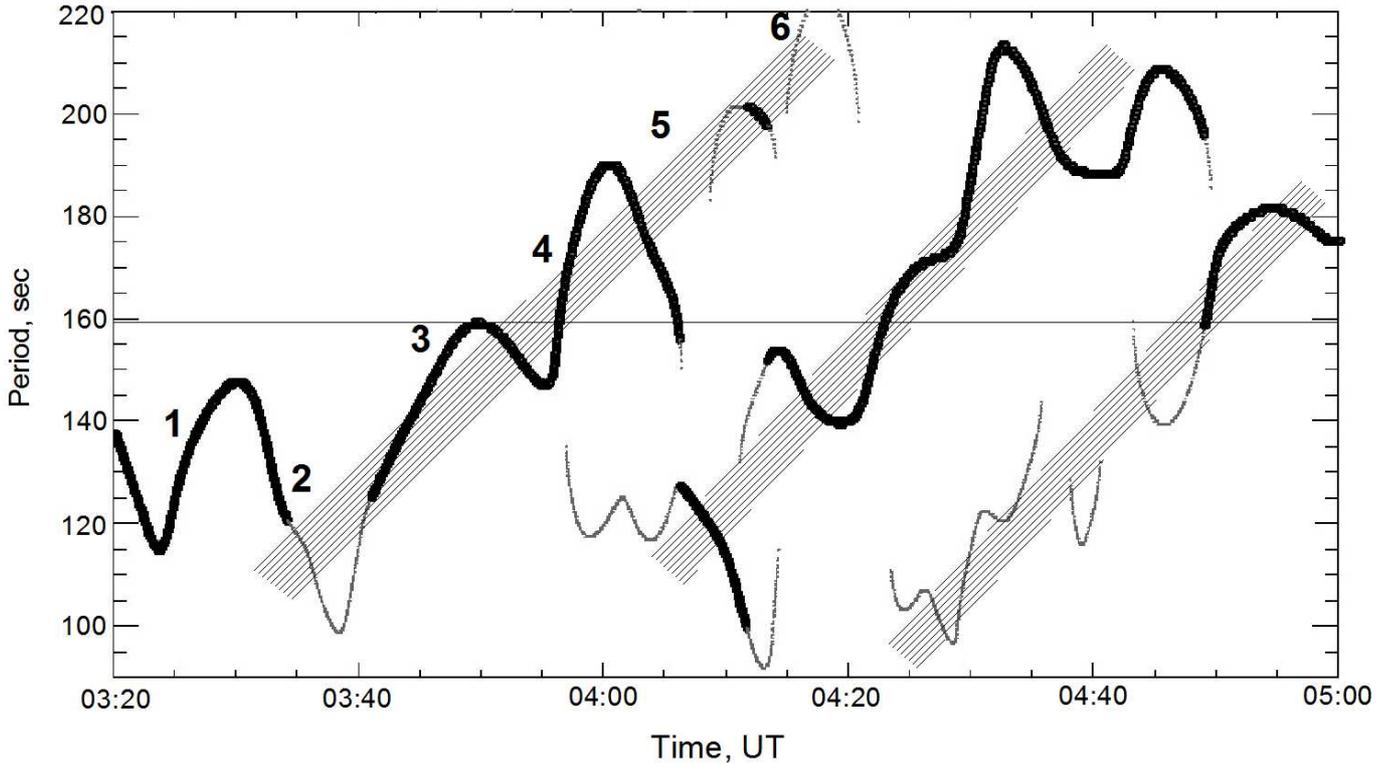}
\caption{Wavelet skeleton of the NoRH correlation curve in the 90-220 s period band as a function of time, obtained at 17 GHz
in the polarisation channel on May 04, 2005 03:20 - 05:00 UT. The numbers indicate the times of the narrowband images shown in Fig.~\ref{fig6}. The mesh strips indicate the prevailing trend of drifts. }
\label{fig5}
\end{figure*}

We built a power wavelet spectrum of the correlation curve and extracted the periods about 3-min, constructing a narrowband signal and wavelet skeleton shown in Figure~\ref{fig4}.
It is evident that the low-frequency amplitude modulation of the 3-min oscillation has a form of repetitive oscillation trains of varying amplitude.  To obtain parameters of this modulation and visualise the power distribution of the oscillations on 
the period-time plane, the power wavelet spectrum was made for the narrowband signal filtered in the interval of 120 to 240 s periods.  The wavelet spectrum is shown in Fig.~\ref{fig4}b. Shaded contours show the power distribution normalised to the maximum value. The magnitude contours range from 0.2 to 0.9.  Darker areas represent more powerful oscillations. Irregular distribution of the power of 3-min oscillations in time and periods is clearly seen.
The irregularity in the power is also seen in the main periods of the oscillations, asymmetry of the wavelet spectral contours of the individual trains, and also in the variations of the train duration and time intervals between them. 
We integrate the wavelet coefficients over the frequency band, getting the time profile of the 3-min oscillation power (Fig.~\ref{fig4}c. A significant periodicity in the train modulation is clearly seen. In the analysed time interval, we find about twelve trains of different maximum power and duration.  The train numbers are indicated on the their tops. 

We would like to point out that the periods evident in GWS (Fig.~\ref{fig3}) correspond to the characteristic times of the wave trains seeing in  Fig.~\ref{fig4}c. To verify this observation, we found the time distance between the beginning and ends of each oscillation train. The duration of individual trains range between  7 and 21 min with the average of about 13 min.  These values coincide with the evident peaks in the GWS spectrum in the range from 8 to 16 min.  Therefore, it can be asserted that the average duration of 3-min trains is about 13 minutes with variation close to 8, 12 and 16 min. Moreover, the long-period oscillations with the periods of 30 and 50 min are possibly related to the characteristic time interval of the train occurrence. This possible association has also been pointed out in Chorley et al. (\cite{2010A&A...513A..27C}). 

The observed amplitude modulation of 3-min oscillations is accompanied by the change of the frequency.
These changes can be visually detected by the behaviour of the power distribution in the individual trains in the frequency range (panel b, Fig.~\ref{fig4}). To study this effect, we constructed the wavelet skeleton of the NoRH correlation curve (Fig.~\ref{fig4}d).  This allows us to track the time and frequency dynamics of both, global and local extrema, and to obtain the values of frequency drifts for different phases of the sunspot activity.  We can see that the period of 3-min component is not stable during the observations.  There are repetitive frequency drifts with the rate from 3 to 5 mHz/hour, mostly occurring in the frequency band of 8.3-4.2 mHz, both increasing and decreasing the frequency.  In the analysed time interval,
negative drifts prevail, e.g. for trains 2, 3, 7, 8, 10 and 11. There are also intervals with positive drifts (trains 5 and 12), and with variable drifts (trains 1, 4, 6 and 9).  The signs of the drifts are indicated in circles in the figure. 

The times of the beginning and ends of the individual trains are shown by broken curves in Fig.~\ref{fig4}d.  Comparing the times of the power peaks of the 3-min oscillation trains (Fig.~\ref{fig4}c), and their start and end times with the temporal dynamics of the drift skeletons, we see that the beginnings of trains and the frequency drifts coincide in the majority of cases.  
The period of 172 s, corresponding to the maximum of the spectral power, determined by GWS, usually falls in the middle drift (indicated by the arrows in Fig.~\ref{fig4}d). The drift typically begins from shorter periods, of about 140 s, and ends at about 220 s. The formation of another train, with the frequency significantly different from the previous one, is possible before the end of the previous train. Trains 7, 8 and 9 trains in 02:05-02:20 UT are good examples.
In individual trains, the linear, systematic frequency drift can be accompanied with some deviations, e.g. at 01:47-02:15 UT and 02:27-02:55 UT.  

Similar results were found for other single sunspots analysed (02-05.05.2005 NOAA 10756, 23-25.01.2002 NOAA 9787, 04.07.2006 NOAA 10898, 15.08.2006 NOAA 2006 and 29.4.2007 NOAA 10953): 3-min oscillations are subject to the persistent modulation in all analysed observations.  The modulation depth varies in time by two orders of magnitude. Wavelet skeletons show both positive and negative frequency drifts. 
Statistical properties of the parameters of 3-min oscillations determined by the ten events are shown in Table 1. 

\subsection{Spatial structure of  3-min radio oscillations}
%

\begin{figure*}
\centering
\includegraphics[width=15cm, height=11cm]{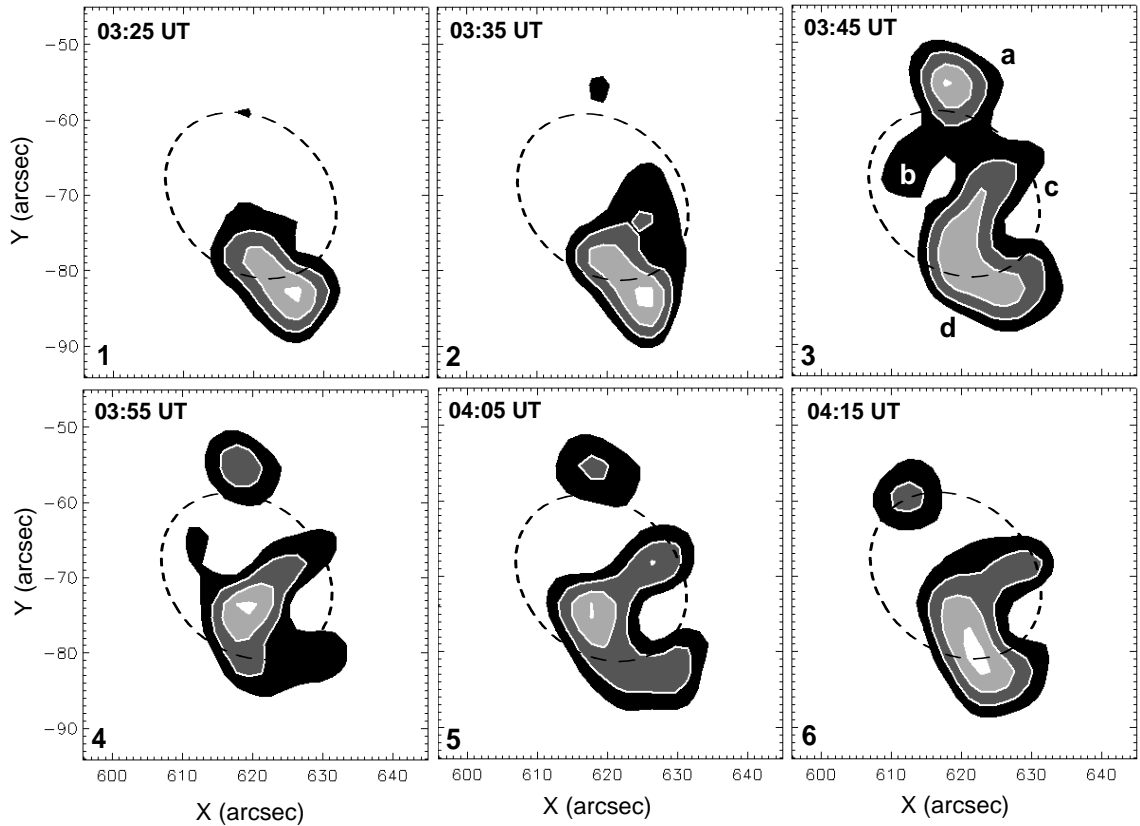}
\caption{Time evolution of the spatial sources of 3-min oscillations of  the microwave polarisation signal from sunspot NOAA 10756, obtained with NoRH at 17 GHz on 2005 May 04, during the 3-min frequency drift 03:20-04:30 UT. The grey colour scale shows the power of the oscillation. The dashed curve indicates the umbra-penumbra boundary. The integer numbers label the instants of time shown in Fig.~\ref{fig5}. The letters label individual sources of the oscillations.}
\label{fig6}
\end{figure*}

To study the relationship between the frequency drift and spatial changes in sunspot-associated radio sources, we consider sunspot NOAA 10756 during 3:20-05:00 UT on May 4, 2005. The GWS of the NoRH correlation curve in the 17 GHz polarisation channel shows well-pronounced 3-min oscillations. The wavelet skeleton with the set of global and local extrema of the wavelet coefficients is shown in  Figure~\ref{fig5}. In the analysed time interval, there are three well-seen negative frequency drifts, similar to the detected on the previous day (Fig.~\ref{fig4}). 
The drift speed is about 7 mHz/hour.  The typical duration of the drift is 40 min. The start and end times of consecutive trains are seen to overlap.
For the times, indicated in Fig.~\ref{fig5}, we constructed narrowband images of the spatial structure of the 3-min oscillation sources with the use of the PWF technique, shown in Figure~\ref{fig6}.  
It is evident that there are significant spatial changes in the narrowband 3-min oscillation sources during 
the development of the drift: new drifts are clearly associated with the appearance of new sources of the oscillations.  
For example the beginning of the drift at about 3:40 UT is accompanied by the
development of a new source labelled by (a) in addition to the already existing source (d) that corresponds to the previous
act of the frequency drift. 

\subsection{Spatial structure of 3-min oscillation trains in EUV}
%

In order to confirm the results of the frequency-spatial changes in 3-min oscillations, determined at the transition  region 
in the microwave band, we analyse the data obtained at other levels of the sunspot atmosphere, in the UV and EUV bands
on May 4, 2005 using TRACE, and on December 8, 2010 using the SDO/AIA.

\subsubsection{The event on May 04, 2005 observed with TRACE}
\label{trace_an}

\begin{figure*}
\centering
\includegraphics[width=17cm, height=8cm]{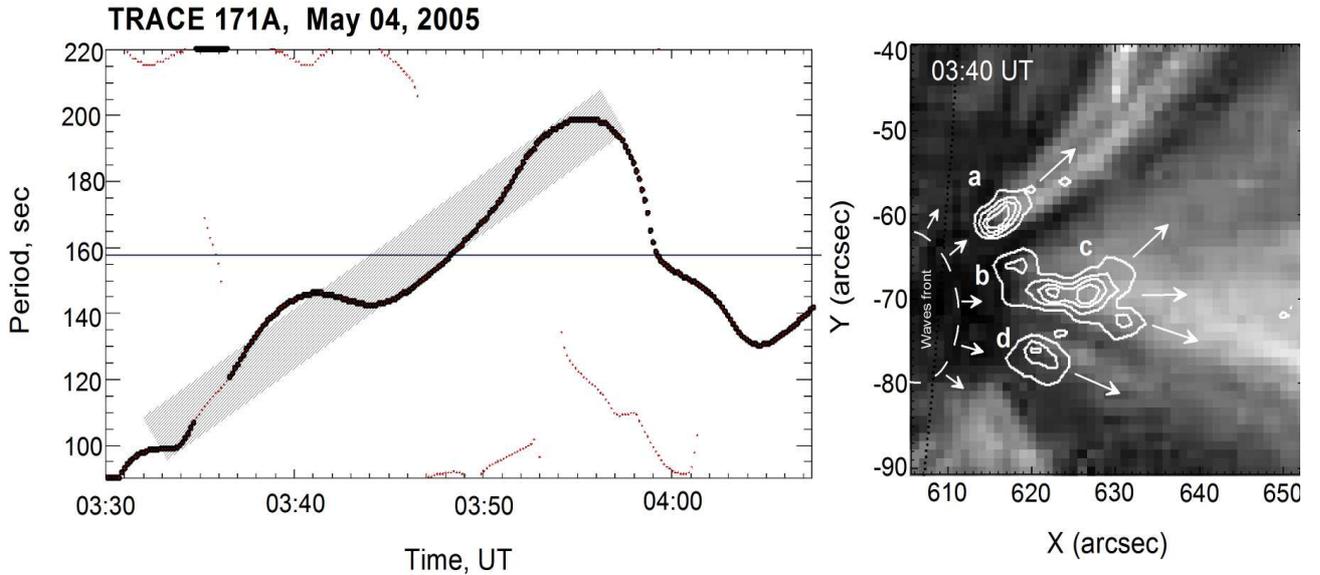}
\caption{Left panel: Wavelet skeleton of 3-min oscillations observed in the EUV emission over sunspot NOAA 10756 on May 04, 2005 at 03:30-04:08 UT. The negative trend of the frequency drift is shown by the mesh strip. Thick lines correspond to the global maxima of wavelet coefficients, and the thin lines to the local ones. Right panel: Overlap of the narrowband image of 3-min EUV oscillation  sources (shown by white contours) on the integral EUV image at 03:40 UT. The letters a, b, c and d correspond to the fine 3-min oscillation radio sources in Fig.~\ref{fig6} at 03:45 UT. The arrows indicate the apparent direction of the wave propagation.}
\label{fig7}
\end{figure*}

EUV images of the active region NOAA 10756 with the cadence of about 9 s were obtained with TRACE at 171\AA\  on May 4, 2005 at 03:30-04:10 UT. After the standard calibration and pre-processing procedure, a data cube of the 50x50 arcsec
images of the active region was obtained. A movie of the data cube clearly shows periodic EUV disturbances propagating along the plasma fan from the sunspot outwards. Integrating the data cube over the spatial coordinates we obtain the time profile of the EUV flux from the active region. The wavelet skeleton constructed in the 90-220 s period band shows a negative frequency drift of the 3-min oscillations (Fig.~\ref{fig7}, left panel), similar to the drift detected in the microwave in the transition region (Fig.~\ref{fig5}). However, the drift speed is about 13 mHz/hour, which is almost two times higher than detected at 17 GHz.

We used the PWF-method investigating the spatial structure of sources of 3-min oscillations at different heights 
and their relationship to the frequency drift, at the time about 03:40 UT, when significant changes 
were detected in the microwave band (Fig.~\ref{fig7}, right panel). It is evident that the power of 3-min waves 
concentrates in some open plasma structures (which were called \lq\lq wave traces" in Sych et al. \cite{Sych2009,Sych2010}).
Although there is almost an order  of magnitude difference in the spatial resolution of NoRH and TRACE,
the 3-min oscillation sources determined in EUV have similar spatial location and shape as the sources determined in microwaves (see the details labelled by the letters a, b, c and d  in Fig.~\ref{fig6} and \ref{fig7}, right panel). 
Like in the microwave data, the appearance of new sources of 3-min oscillations coincides with the beginning of a 30-min train as well as with the frequency drift. 

 \begin{figure*}
\centering
\includegraphics[width=15cm, height=11cm]{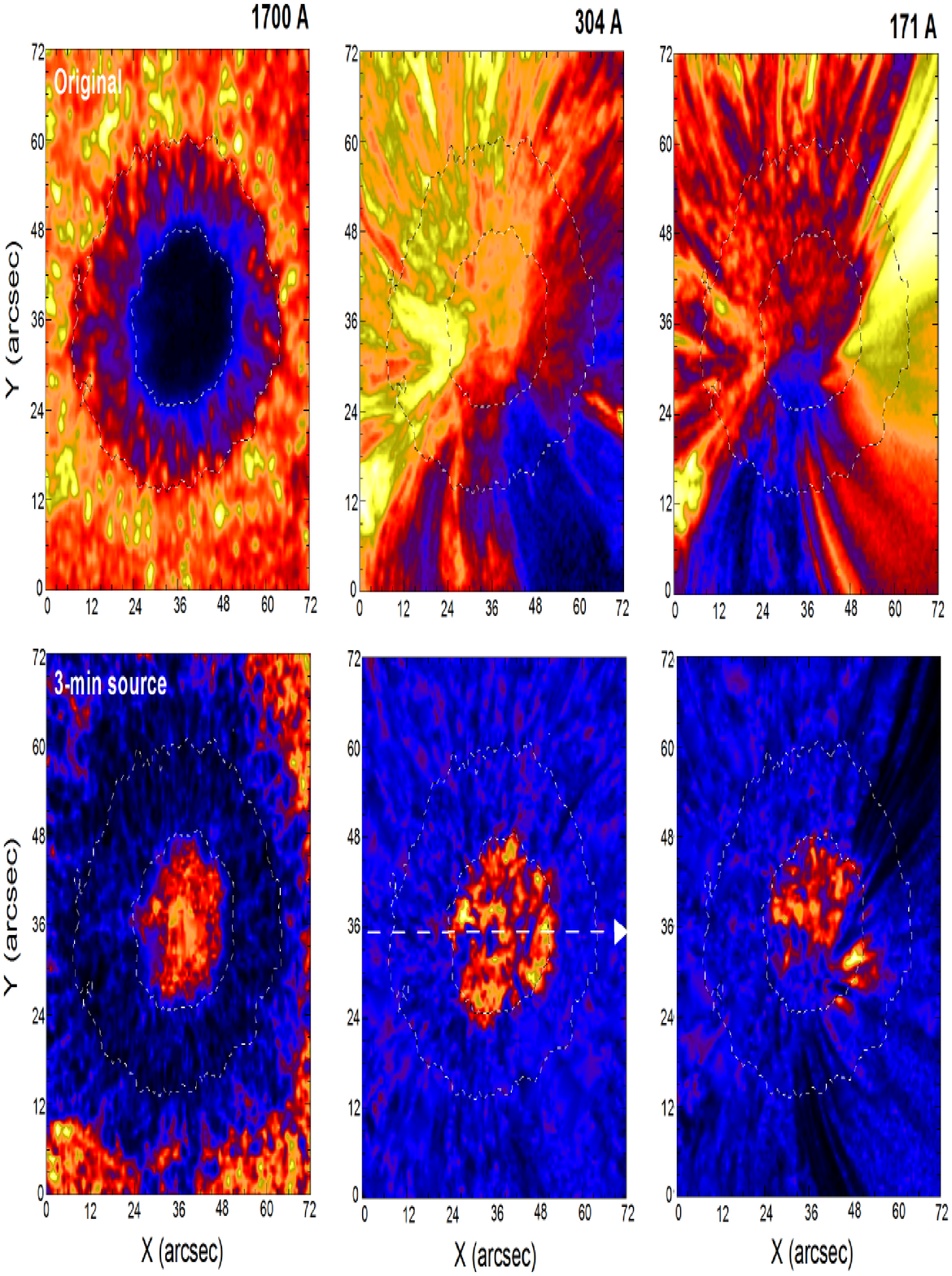}
\caption{Top panels: Images of NOAA 11131 at 1700 \AA\ (photosphere), 304 \AA\ (chromosphere) and 171 \AA\ (corona) on December 8, 2010 at 02:30 UT. Bottom panels: spatial distribution of 3-min oscillations, obtained by filtering  in the 5.2-9.5 mHz band. Thin dotted lines indicate the umbra-penumbra boundary at the photospheric level. The thick dashed line in the narrowband image at 304 \AA\ shows the slit of the time-distance plot.}
\label{fig8}
\end{figure*}

\subsubsection{The event on December 08, 2010  observed with SDO/AIA}

Active region NOAA 11131 was observed by SDO/AIA on December 8, 2010. Sequences of 72x72 arcsec images 
were obtained with the 24-s cadence over the time interval 02:30-03:30 UT at 1600, 1700, 171, 304, 193, 211, 131, 335 and 94 \AA\ wavelengths. This set of observational wavelengths allows us to study properties of 3-min oscillations, in particular the frequency drifts, 
from the photospheric up to coronal heights (Figure~\ref{fig8}, top panel). 
Applying PWF, we make data cubes of the spatial distribution of narrowband 3-min oscillations. 
Snapshots of the spatial structure of the oscillations are shown in the bottom panels of Fig.~\ref{fig8}.

We see that sources of 3-min oscillations are mainly concentrated in the umbra. For lower layers of solar atmosphere, i.e. at 1700 \AA , the sources have the shape of symmetric patches that fill almost uniformly the central part of the umbra. In the chromospheric level, i.e. at 304 \AA, we see the fragmentation of the oscillation source that gains a spiral shape, filling up
the whole umbra. In the coronal level (171 \AA), further deformation of the oscillation source takes place: a part of the source has a radial sector structure, while the other part is stretched along coronal plasma loops that form the active region fan. A similar shape of the oscillation source is seen in higher-temperature emission lines, but the power of 3-min oscillations in this emission is significantly lower. 

Information about frequency drifts and wave propagation can be obtained from time-distance plots, constructed
for the time variation of the emission intensity along a certain one-dimensional spatial slit.
For example, such a plot
for a horizontal slit passing through the centre of the sunspot in the 304\AA\ data cube 
is shown in Figure~\ref{fig10}.  
The wave structure of 3-min oscillations, propagating symmetrically from the sunspot centre, the consequent decrease in the propagation speed and damping at the umbra-penumbra boundary, are clearly seen. 
 Analysis of the time characteristics of the signals was performed by the same fashion as of the TRACE data
(Sec.~\ref{trace_an}) separately for each observational wavelength. 

\begin{figure*}
\centering
\includegraphics[width=17cm, height=10cm]{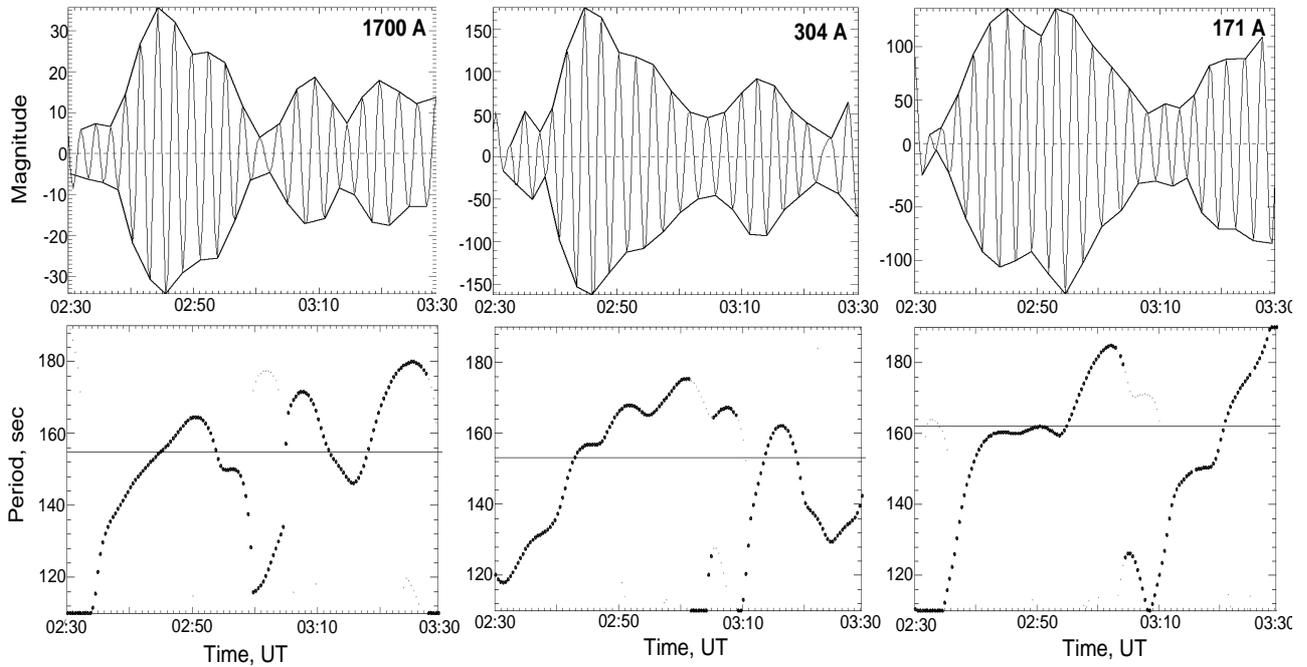}
\caption{Top panels: Time profiles of 3-min oscillation trains in the emission intensity of the 1700, 304 and 171\AA\ 
signals obtained with SDO/AIA on December 8, 2010 at 2:30-03:30 UT. Envelopes of the trains are shown by 
the thick curves. Bottom panels: wavelet skeletons of the signals. The continuous line shows the period corresponding
to the maximum spectral power.}
\label{fig9}
\end{figure*}

\begin{figure}
\centering
\resizebox{\hsize}{!}{\includegraphics[viewport=6 252 557 551]{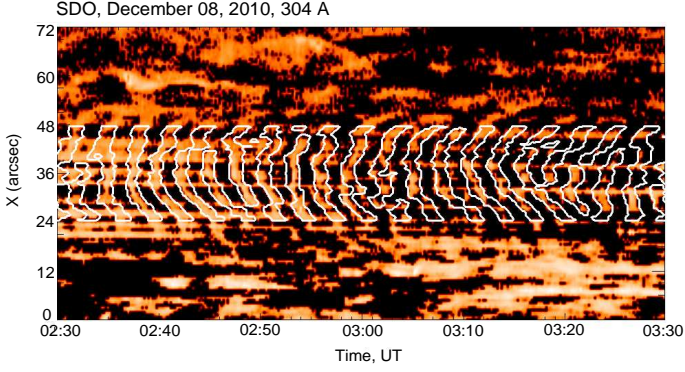}}
\caption{The time-distance plot of the active region NOAA 11131 on December 8, 2010 obtained by SDO/AIA at 304 \AA\ . The intensity oscillations are presented in the logarithmic scale. The white contours show the narrowband wave fronts 
in the 5.2-9.3 mHz band. The spatial slit is positioned across the sunspot, through its centre, and is shown in Fig.~\ref{fig8}.}
\label{fig10}
\end{figure}

Figure ~\ref{fig9} shows the amplitude and phase modulation of 3-min oscillations 
for three wavelengths, 1700, 304 and 171 \AA\ , which correspond to three different heights, and their 
wavelet skeletons.  As in our previous findings, the oscillation trains last typically from 8 to 20 minutes (with the average of about 12 min),
showing partial overlap. Times between individual trains are about 30 minutes. Wavelet skeletons show that
the rising phase of the trains is accompanied with negative frequency trains.
The average period of 3-min oscillations is 155 s at 1700 \AA,  153 s at 304 \AA\ and
 162 sec at 171 \AA\ . Negative frequency drifts dominate in all observational channels. The drift speed at the photospheric level (1600 \AA, 1700 \AA) is 
 about 4-5 mHz/hour, at the chromospheric level (304 \AA) about 5-7 mHz/hour, and at microwaves (NoRH, 17GHz) 5.6-8 mHz/hour and in the corona (171, 193, 211 and 131 \AA) about 11-13 mHz/hour. Thus, the absolute value of the drift speed increases by almost two times during the propagating from the 
sub-photospheric levels to the corona. In higher temperature emission (335 and 94 \AA) the 
3-min signal is weak, and hence it is impossible to get reliable information about the drifts. 

\section{Discussion}

Using high-quality EUV and microwave data obtained with SDO/AIA, TRACE and NoRH  and modern analytical techniques (PWF-method and wavelet skeletons), we studied complex temporal dynamics of 3-min oscillations above sunspots. The dynamics includes variations in height, spatial, temporal and frequency structure of the oscillations.  
Spatially-integrated signals obtained at different heights and with different instruments show common behaviour:
the oscillations experience amplitude and frequency modulation, which is accompanied with the change of the
spatial sources. The amplitude modulation has the form of wave trains, of the duration from
7 and 21 min  and the repetition period of about 30-60 min. The frequency modulation has the form of frequency 
(or period) drifts in the 2-4 min period range. Typically, the beginning and end times of the drifts coincide with the
trains of the oscillations, which indicates their phenomenological relationship.
Mainly negative drifts (the gradual increase in the period of 3-min oscillations) are observed. 
Sometimes two drifts co-exist, i.e. during the end of the previous drift,
when the frequency is in the low-frequency part of the 3-min spectrum, near 200 s, a new drift appears near 160 s.
Development of new 3-min oscillation trains is typically accompanied with the appearance of new spatial sources
(the regions of the enhanced power of 3-min oscillations in the horizontal plane) in the sunspot atmosphere. 
In the corona, spatial sources of 3-min oscillations develop along certain
magnetic flux tubes of the magnetic fans above the sunspot. Narrowband time-distance plots show that in 
those structures 3-min oscillations have a form of outwardly propagating waves. This behaviour is well
consistent with recent numerical modelling (Botha et al. \cite{2011ApJ...728...84B}) and the analysis
of imaging and spectral EUV data (e.g. De Moortel \cite{2009SSRv..149...65D}; Verwichte et al. \cite{2010ApJ...724L.194V}). 
Also, there are coronal sources of another type: compact, of a symmetric, patch-like shape, which can perhaps be associated with the magnetic flux tubes with a narrow angle to the line-of-sight. 

\subsection{A possible mechanism for frequency drifts}

One of the possible mechanisms explaining the observed properties of 3-min oscillations oscillation is the response of 
the sharply stratified atmosphere to pulses coming from lower regions. The pulses can be excited by magnetic reconnection, granulation or other explosive or impulsive events taking place in sub-photospheric, photospheric or lower chromospheric regions. 

Consider a gravitationally stratified plane atmosphere with a vertical magnetic field. We assume that $p_0(z)$ and $\rho_0(z)$ are the plasma
pressure and density, respectively; and the temperature, $T_0$, is constant.  
Then sound speed, $c_s=\sqrt{\gamma p_0/\rho_0}$, is also constant. The pressure scale height is 
\begin{equation}\label{H}
H={{k_B T_0}\over {{\hat m}g}}, \end{equation} 
where $k_B$ is the Boltzmann constant, $g$ is the gravitational acceleration and ${\hat m}$ is the mean molecular weight. This model of the solar lower atmosphere
is obviously over-simplified, but we use it for illustration purposes only.

Propagation of linear sound waves in the stratified medium is governed by the Klein-Gordon equation, which implies the 
acoustic cut-off frequency
\begin{equation}\label{cut-off}
\omega_c={{c_s}\over {2 H}}.
\end{equation}
Waves with frequencies higher than $\omega_c$ can propagate upwards, while the lower-frequency waves are evanescent. However, evolution of an initially broadband pulse leads to a formation of an oscillatory wake (Lamb \cite{Lamb}, Rae 
\& Roberts \cite{Rae}):
\begin{equation}\label{V_1}
u_1=u_0J_0\left (\sqrt{\omega^2_c t^2-{{z^2}\over {4
H^2}}}\right ){\cal H}\left (\omega_c t -{{z}\over {2 H}}\right
)\exp{{{z}\over {2 H}}},
\end{equation}
where $u_1$ is the vertical component of the velocity, $u_0$ is the initial amplitude, $J_0$ is the Bessel function
of the zeroth order, and ${\cal H}$ is the Heaviside function. The time is measured from the time when the pulse was launched.

As the chromospheric cut-off period is about 3-min then the observed oscillations can be explained as wakes behind the pulses
propagating along the vertical magnetic field (Fleck and Schmitz \cite{Fleck}, Kalkofen et al. \cite{Kalkofen}, Sutmann and Ulmschneider \cite{Sutmann}, Kuridze et al. \cite{Kuridze}).
However, each active region may have a fine structure in the horizontal direction, in the form of flux tubes with different plasma temperatures, as the strong magnetic field prohibits thermal conduction across the field. Then, the pulses propagating in different parts of the active region may set up the wakes with different oscillating periods. The radio observations may
not resolve the fine horizontal structure of the oscillations due to insufficient spatial resolution. Then the oscillations with different frequencies will be superimposed, and the resulted observed oscillation may have a complex structure.

\begin{figure*}
\centering
\includegraphics[width=15cm, height=11cm]{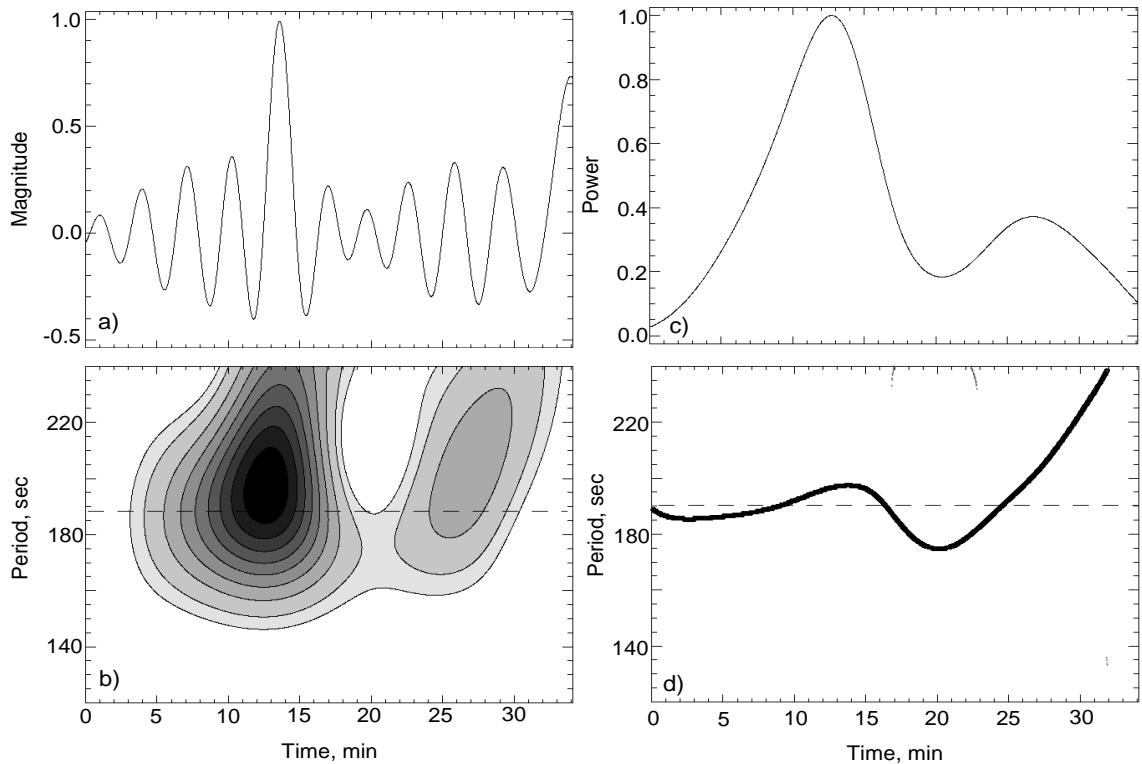}
\caption{A model of the observed behaviour of 3-min oscillations: superposition of two 
oscillatory wakes formed in two different magnetic flux tubes in a sunspot's umbra. Panel (a): the signal; panel (b): its wavelet spectrum; panel (c): time dependence of the oscillation power;
panel (d): wavelet skeleton of the signal.}
\label{fig11}
\end{figure*}

In order to illustrate this scenario, in Figure~\ref{fig11} we plot the linear superposition of two wakes, which could be formed by two acoustic pulses propagating along two different flux tubes in the umbra. The ratio of the
cut-off frequencies in these two regions is taken to be 0.8. The second pulse was launched 810 s later than the first pulse. 
We also show the wavelet spectrum of the signal, time variation of the oscillation power, and the wavelet skeleton with the global maximum. The time interval between the wave trains is about 14 min, which actually is the interval between the consecutive pulses. The oscillation period slightly increases (decreases) before (after) the maximum of the oscillation amplitude. Thus, the behaviour of the modelled oscillation is very similar to the observed behaviour of 3-min oscillation trains. 

\section{Conclusions}

We summarise results of this study as follows:

   \begin{enumerate}

      \item We analysed 3-min oscillations in the NoRH polarisation correlation curves at 17 GHz, obtained for eleven time intervals when there was only one powerful sunspot on the solar disk. The relative amplitude of 3-min oscillations was about 3-8\% reaching 17\%. Spectral analysis demonstrated the presence of significant peaks in the range of 2-4 min with a power concentration near 3 min period. The 3-min oscillations experience amplitude modulation in the form of periodic wave trains, lasting for 8-20 min with 13 min in average.
The repetition rate of the trains is 20-40 min with 30 min in average.
There were also low-frequency peaks in the spectrum, corresponding to the periods of  10-20 min, and 30-60 min. Apparently these long-period oscillations coincide with the characteristic times of the 3-min oscillation trains.
       \item In addition, 3-min oscillations are subject to frequency modulation: their period
varies in the range 90-240 s. During the oscillation train, the frequency modulation has often a form
of a steady frequency drift. The beginning and end of the drift coincide with the beginning and
end of the amplitude modulation. In the majority of cases, the negative drift, when the 3-min oscillation period gradually increases, is seen. In different trains of 3-min oscillations observed in the same sunspot, there can be both positive and negative drifts. The start and end times of consecutive drifts can overlap.
This behaviour is seen at all levels of the sunspot  atmosphere. The speed of the drift is 4-5 mHz/hour in the photosphere, 5-8 mHz/hour in the chromosphere, and 11-13 mHz/hour in the corona. Thus, apparently, the drift speed grows with height. In addition to the gradual drift, the instant frequency of 3-min oscillations experience positive and negative fluctuations.
      \item Analysis of the spatial structure of 3-min sunspot oscillations observed in microwaves and EUV 
   shows that the start of drifts coincides with the appearance of sources
of the oscillations (the regions with the enhanced power of 3-min oscillations).   
	\item Observed spatial-frequency-temporal properties of 3-min oscillations 
can be interpreted in terms of two simultaneously-operating mechanisms. The first effect is connected to the negative frequency drift due to the dispersive evolution of upward propagating sub-photospheric pulses, guided by the sunspot magnetic field. The repetition rate of the pulses (30-60 min) is similar to the long-period oscillations of sunspots, which suggest their possible association.  The second effect is the spatial separation of the propagating pulse
by several individual magnetic flux tubes, of different curvature, temperature and line-of-sight
angles. In particular, the frequency drifts may take place due to the changes in the tube curvature with height.
  
\end{enumerate}

\begin{acknowledgements}

The work was supported by the grants RFBR 10-02-00153 and 11-02-10000-k. RS
is supported  by the Chinese Academy of Sciences Visiting Professorship for Senior International Scientists No. 2010T2J24. TZ is supported by the Austrian Fonds zur F\"orderung der wissenschaftlichen Forschung (project P21197-N16) and the Georgian National Science Foundation (grant GNSF/ST09/4-310).
The TRACE, SDO and NoRH data were obtained from the TRACE, SDO and NoRH databases. The authors acknowledge the TRACE, SDO and NoRH consortia for operating the instruments and performing the basic data reduction, and especially for the open data policy. Wavelet software was provided by C. Torrence and G. Compo, and is available at URL: http://paos.colorado.edu/research/wavelets/.

\end{acknowledgements}

\end{document}